\UseRawInputEncoding
\documentclass[12pt]{iopart}
\usepackage{graphicx}
\usepackage{overpic}
\usepackage{amssymb}

\begin{document}

\title[]{Polychromatic drivers for inertial fusion energy}

\author{Yao Zhao$^{1,3,5}$, Zhengming Sheng$^{2,3,4}$, Zijian Cui$^{1,5}$, Lei Ren$^{1,5}$, Jianqiang Zhu$^{1,3}$}

\address{$^1$Key Laboratory of High Power Laser and Physics, Shanghai Institute of Optics and Fine Mechanics, Chinese Academy of Sciences, Shanghai 201800, China}
\address{$^2$Key Laboratory for Laser Plasmas (MoE), School of Physics and Astronomy, Shanghai Jiao Tong University, Shanghai 200240, China}
\address{$^3$Collaborative Innovation Center of IFSA (CICIFSA), Shanghai Jiao Tong University, Shanghai 200240, China}
\address{$^4$SUPA, Department of Physics, University of Strathclyde, Glasgow G4 0NG, UK}
\address{$^5$Author to whom any correspondence should be addressed.}

\ead{yaozhao@siom.ac.cn, cuizijian@siom.ac.cn, leir89@siom.ac.cn}

\begin{abstract}
Although tremendous achievements have been made toward inertial confinement fusion, laser plasma instabilities (LPIs) remain to be an inevitable problem for current drive schemes. To mitigate these instabilities, significant efforts have been paid to produce high-power broadband ultraviolet lasers. However, no practical scheme has been demonstrated up to now for efficient triple-frequency conversion of broadband laser. Here we propose the design of polychromatic drivers for the generation of multicolor beams mainly based upon the optical parametric amplification, which can significantly enhance the third-harmonic conversion efficiency. Each polychromatic light has four colors of monochromatic beamlets with a full spectrum width of 3\%, and the beamlet colors of any two adjacent flanges are different. The suppression effects of such polychromatic lights have been investigated via large scale particle-in-cell simulations, which indicate that more than 35\% of the incident energy can be saved from the LPIs compared with monochromatic lasers for the direct-drive scheme, or high-density filled target for the indirect-drive scheme. The proposed polychromatic drivers are based on the matured technologies, and thus may pave the way towards realization of robust and high-efficiency fusion ignition.
\end{abstract}

Accepted by New Journal of Physics (https://doi.org/10.1088/1367-2630/ac608c)

\section{Introduction}

Achieving nuclear fusion energy on the Earth has long been a grand challenge since the idea of laser confinement fusion was proposed in the 1970s \cite{nuckolls1972laser}. Recently, the fusion energy yield of 1.3MJ has been achieved at the National Ignition Facility (NIF) \cite{zylstra2022burning}, which represents a new milestone towards the realization of inertial confinement fusion (ICF). In spite of this, the enhancement of the beam-target coupling efficiency remains a critical issue for achieving high-efficiency ignition. Laser plasma interaction is an inevitable process in ICF \cite{froula2010experimental,Pesme2002Laser,Hinkel2004National}, where a target containing deuterium and tritium fuel is directly (direct drive) or indirectly (indirect drive) compressed by megajoule laser beams \cite{craxton2015direct,Lindl2014Review,Hurricane2014Fuel,Betti2016Inertial,zhang2020enhanced}. Laser plasma instabilities (LPIs) can be fully developed at the $10$ picosecond scale in the interactions of high-power lasers with large scale plasmas \cite{kruer1988physics}. The complex and sensitive LPIs, such as stimulated scattering instabilities and two-plasmon decay (TPD), are among the major obstacles to achieving ignition, as they can cause significant laser energy loss \cite{ping2019enhanced,rosenberg2018origins}, asymmetric and insufficient compression \cite{Igumenshchev2012Crossed,InterplayStrozzi,2012Multistep}, and target preheating \cite{Dewald2010,Smalyuk2008Role,christopherson2021direct}. Stimulated scattering instabilities including stimulated Raman scattering (SRS) and stimulated Brillouin scattering (SBS) are the decay of incident lasers into scattered lights via induced plasma density perturbations, which cause considerable incident energy loss and produce a large number of hot electrons \cite{montgomery2016two}. As an example for indirect drive, the minimum laser energy required for ignition scale is inversely proportional to the hohlraum-to-capsule coupling efficiency \cite{herrmann2001ignition,MtV2004}. Therefore, mitigation of LPIs will contribute to the high-efficiency and robust ignition performance. A few ideas have been proposed to inhibit LPIs, such as laser smoothing techniques \cite{skupsky1989improved,lehmberg1983use,moody2001backscatter}, broadband lasers \cite{thomson1974effects}, a spike train of uneven duration and delay \cite{albright2014control}, and near-vacuum hohlraum \cite{hopkins2015first}, etc.

Suppression of LPIs via broadband lasers has been investigated from the early 1970s \cite{thomson1974effects,Estabrook}. Both the theoretical model and numerical simulation indicate that the linear growth of LPIs can be reduced when the laser bandwidth is much larger than the growth rate \cite{Santos001,zhao2015effects}. Therefore, typically the bandwidth level $\gtrsim1\%\omega_0$ ($\gtrsim3.51$nm for the triple-frequency lasers) is required for the LPI inhibition, where $\omega_0$ is the central frequency. However, similar to the control of LPIs, the phase matching conditions required for the frequency conversion of broadband lasers are also hard to be satisfied \cite{cui-01,cui-02,cui-03}. As a result, the bandwidth of the available ultraviolet laser is extremely limited, and the conversion efficiency is greatly reduced \cite{cui-03,cui-04}. Significant efforts have been paid to produce the ultraviolet broadband lasers with high efficiency \cite{Obenschain,Weaver}. However, no practical scheme has been demonstrated up to now for efficient triple-frequency conversion of broadband lasers \cite{dorrer2021broadband}.

Different from conventional broadband lasers, in this article we propose the design of polychromatic lights with each beam composed of multiple colors of monochromatic beamlets propagating in the same direction. The polychromatic driver generates wide spectrum multicolor lights based upon the optical parametric amplification (OPA), and each monochromatic beamlet is frequency converted independently, which skirts the low conversion efficiency of the broadband triple frequency. The LPIs developed by the polychromatic lights have been studied via theory and simulations, which show better inhibition effects than continuous broadband lasers at a long time scale \cite{YaoZ2017Effective,follett2018suppressing}. In the following, a feasible design of the optical system is proposed based on the theory model of LPIs driven by polychromatic lights. Finally, the enhancement of beam-target coupling is validated via large-scale simulations regarding to the design parameters.

\section{Theoretical model for LPI inhibition via polychromatic light}

The fundamental requirements for the development of LPIs are the wave-vector matching condition $\vec{k}_0=\vec{k}_p+\vec{k}_s$ and the frequency matching condition $\omega_0=\omega_p+\omega_s$, where $\vec{k}_0$, $\vec{k}_p$ and $\vec{k}_s$ are the wave-vectors of incident light, plasma waves (including electron plasma wave and ion acoustic wave) and scattering light, and $\omega_p$ and $\omega_s$ are the frequencies of plasma waves and scattering light, respectively. The frequency difference detunes the phase matching of the three-wave coupling. However, there is an applicable finite bandwidth for the satisfaction of the phase matching conditions, which determines the decoupling threshold of the beamlets. When the bandwidth is much smaller than the decoupling threshold, the components are strongly coupled to develop LPIs. The low-coherent laser with a continuous spectrum may behave badly in the convective process and the nonlinear wave-particle interactions, where the amplitude modulation and the extended interaction region will compensate the reduction of temporal growth rate \cite{zhao2015effects,zhao2017stimulated,guzdar1991effect}. Therefore, decoupling conditions are critical to the design of polychromatic lights. Both the single-beam and collective LPIs can be effectively mitigated, as long as the amplitude of each decoupled beamlet dissatisfies the drive threshold.

\begin{figure}
\centering
    \begin{tabular}{lc}
        \begin{overpic}[width=1\textwidth]{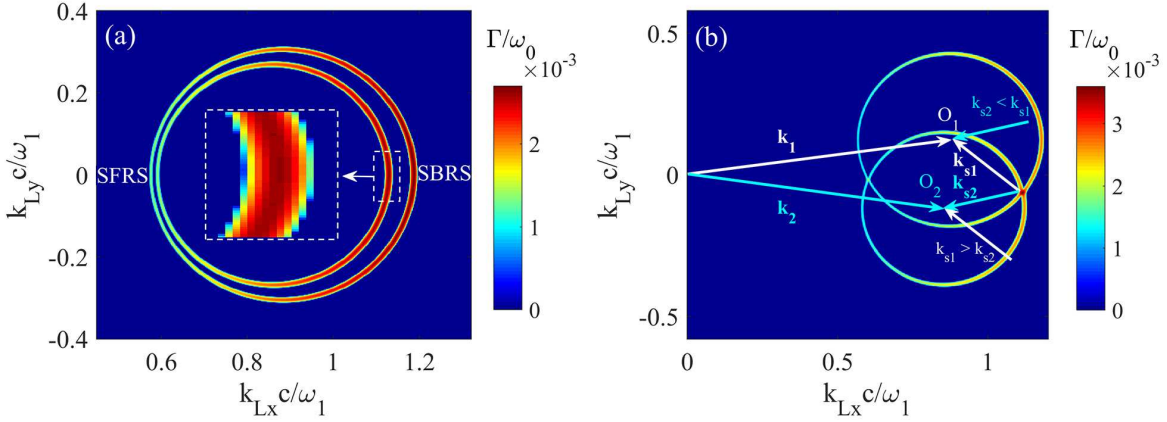}
        \end{overpic}
    \end{tabular}
\caption{Distributions of the SRS growth rate $\Gamma$ driven by two beams in the wavenumber space. The plasma density is $n_e=0.2n_{c1}$, and the electron temperature is $T_e=2$keV, where $n_{c1}$ is the critical density for beam 1. The amplitude of each beam is $a_1=a_2=0.01$ with frequency difference $\delta\omega_0=\omega_1-\omega_2=2\%\omega_1$. (a) The two beams propagate in the same direction, where the locations for stimulated forward (SFRS) and backward (SBRS) Raman scattering are shown. (b) The angle between the two beams is 16$^\circ$.
    }
\end{figure}

The decoupling conditions are obtained based on the two-color LPI model, where the frequency difference between the pumps is $\delta\omega_0=\omega_1-\omega_2$. We first consider the beamlet decoupling threshold of a two-color laser interaction with homogeneous plasmas. An example for SRS is shown in detail \cite{YaoZ2017Effective,zhao2021mitigation}, whose fluid equations are
\begin{equation}
\left(\partial_{tt}-3v_{th}^2\nabla^2+\omega_{pe}^2\right)\tilde{n}_{e}=\frac{\omega_{pe}^2}{4\pi ec}\nabla^2(\vec{\tilde{A}}_s\cdot\vec{a}_{0}),  \label{as}
\end{equation}
\begin{equation}
\left(\partial_{tt}-c^2\nabla^2+\omega_{pe}^2\right)\vec{\tilde{A}}_s=-4\pi ec\tilde{n}_{e}\vec{a}_{0},  \label{ne}
\end{equation}
where $\vec{a}_0=\vec{a}_1+\vec{a}_2$ is the amplitude of incident light, $v_{th}$ is the electron thermal velocity, $\omega_{pe}$ is the electron plasma frequency, $\vec{\tilde{A}}_s=\vec{\tilde{A}}_{s1}+\vec{\tilde{A}}_{s2}$ and $\tilde{n}_{e}$ are the perturbations of scattering light and electron density, respectively. The relation between normalized amplitude and laser intensity is $a_0=\sqrt{I_0 (\mathrm{W}/\mathrm{cm}^2)[\lambda(\mu \mathrm{m})]^2/1.37\times 10^{18}}$. In the strongly coupled regime, the scattering light developed by one of the incident lights will be shared by the other pump beams. Without loss of generality, the frequency coupling term of $\vec{a}_1\cdot\vec{\tilde{A}}_{s2}$ can be written as $\cos[(\omega_1-\omega_{s2})t]=\cos[(\delta\omega_0+\omega_L)t]$, which can resonate with the Langmuir wave with frequency $\omega_L=\sqrt{\omega_{pe}^2+3k_L^2v_{th}^2}$ and wavenumber $k_L$ at $1/\delta\omega_0\gg 1/\Gamma$. In the decoupling regime, the scattering lights of different-color beamlets are independent from each other, and the dispersion relation can be obtained from the Fourier analysis of Eqs. (1) and (2),
\begin{equation}
\omega^2-\omega_{pe}^2-3k^2v_{th}^2=\frac{\omega_{pe}^2k^2c^2}{4}\sum_i^2a_i^2\left(\frac{1}{D_{ei-}}+\frac{1}{D_{ei+}}\right),
\end{equation}
where $D_{ei\pm}=(\omega\pm\omega_i)^2-(\vec{k}\pm\vec{k}_i)^2c^2-\omega_{pe}^2$, $\omega_i$, $\vec{k}_i$ and $a_i$ are the frequency, wavevector and amplitude of $i$-th beam, respectively. The solutions of Eq. (3) regarding to a polychromatic light with two-color beamlets are shown in Fig. 1(a), where the two beamlets propagate in the same direction. One notes that there is a finite width of the dispersion relation along the circle, i.e.,
\begin{equation}
\Delta k_{Li}=a_ik_{Li}\sqrt{\frac{\omega_{pe}(\omega_i-\omega_{pe})}{\omega_i^2-2\omega_i\omega_{pe}}},
\end{equation}
where $k_{Li}$ is the wavenumber of Langmuir wave driven by $i$-th beam with $i=1$ or $2$ here \cite{YaoZ2017Effective}. According to Fig. 1(a), the two beamlets are decoupled only when their instability regions are separated from each other. Even though the width of instability region $\Delta k_{Li}$ is gradually reduced by the decrease of $k_{Li}$, the wavevector of incident light (i.e., the center of the circle) is also shifted to the left $\vec{k}_i(1-\omega_i\delta\omega_0/k_i^2c^2)$ by $\delta\omega_0$. The forward SRS is decoupled when the reduction of the circle radius $\delta r_i=(\omega_i-\omega_{pe})\delta\omega_0/c\sqrt{\omega_i^2-2\omega_i\omega_{pe}}$ satisfies $\delta r_i-\omega_i\delta\omega_0/k_ic^2>\Delta k_{Li}$, which indicates a larger decoupling threshold than the backward SRS $\delta\omega_0/\omega_0\gtrsim3\Gamma_i/\omega_0$ with $\Gamma_i$ being the growth rate of SRS backscattering for the $i$-th beam \cite{YaoZ2017Effective}. Therefore, forward SRS is less sensitive to a small bandwidth than the backward SRS \cite{Santos001,zhao2021mitigation}. However, a frequency difference $\delta\omega_0\sim10^{-2}\omega_0$ is sufficient to decouple the forward modes under $a_0\lesssim0.01$ according to Fig. 1(a). In the linear growth stage, the plasma instability level driven by the decoupled beamlets can be estimated as $\Delta n_{i}=N\delta n\exp(\Gamma_it)$, where $N$ is the beamlet number and $\delta n$ is the initial perturbation. Analogously, the instability level developed by a single monochromatic beam is $\Delta n_{0}=\delta n\exp(\sqrt{N}\Gamma_it)$ in the linear stage, where the beamlet amplitude is $a_i=a_0/\sqrt{N}$ for keeping the same energy with the monochromatic laser. One always finds $\Delta n_{i}/\Delta n_{0}>1$ at the time $t<\ln(N)/[(\sqrt{N}-1)\Gamma_i]<1.7/\Gamma_i$, i.e., the instability level of decoupled beamlets is higher than that of one beam if the latter is saturated long before $1.7/\Gamma_i$ but the latter still grows. Therefore, there exists a maximum beamlet number $N$ to mitigate LPIs in the fluid regime. However, the decoupled beamlets are insufficient to develop LPIs when the amplitude of beamlet is less than the driven threshold in the strongly damped plasmas \cite{Liu2019}.

Equation (3) is also applicable to the multibeam configuration when the lasers develop their respective scattering lights independently \cite{zhao2021mitigation}. The scattering light of one pump will not be shared by the other pump when the wavenumber difference of scattering lights $\delta k_s=k_{s1}-k_{s2}$ is larger than the corresponding instability region width $\Delta k_{Li}$ as shown in Fig. 1(b). The decoupling threshold of SBS can also be obtained from the SBS dispersion relation, which is always less than that of SRS owing to the lower growth rate of SBS \cite{YaoZ2017Effective}.

Now we consider the single-beam convective instability in a heavy damped inhomogeneous plasma with density profile $n_e=n_0(1+x/L)$ based on the Rosenbluth model, where the linear wavenumber mismatch is considered in the phase \cite{rosenbluth1972}. Here $L$ is the density scale length and $x$ is the longitudinal axis, both of them are normalized by light wavelength $\lambda$ in the following. The frequency difference $\delta\omega_0\sim10^{-2}\omega_0$ only introduces a small group-velocity difference of the scattering lights $\delta v_s$, which is therefore ignored. As example for SRS scattering light, one has $\delta v_s/v_s\sim10^{-3}$ at $n_e=0.1n_c$, where $n_c$ is the critical density. When the frequency difference is much larger than the corresponding growth rate, i.e., the temporal growth rate is reduced, the saturation coefficient of convective SRS developed by a two-color beam is \cite{Zhao2019,liu1974raman}
\begin{equation}
G_{\mathrm{SRS}}\approx\frac{2\pi\Gamma_m^2}{v_sv_pK_0'}\left[1+\cos\left(\frac{4L\omega_0\nu_p^2\omega_L^2}{n_0c(\omega_0^2-\omega_{pe}^2)^{3/2}}\delta\omega_0\right)\right],
\end{equation}
where $v_s$, $v_p$ and $\nu_p$ are the group velocities of scattering light and plasma wave, and the damping of the plasma wave, respectively. The modified growth rate can be estimated as $\Gamma_m\approx\Gamma^2/\delta\omega_0$, and $K_0=k_0-k_s-k_L$ denotes the wavenumber mismatch with $K_0'=dK_0/dx$. Based on Eq. (5), we know that the saturation level of electron plasma wave is reduced by the frequency difference $\delta\omega_0\ne0$. An example is considered for a plasma with $n_0=0.08n_c$, $T_e=3$keV and $L=3000\lambda$, and the decoupling condition is approximately $\delta\omega_0\approx2.4\%\omega_0$. Therefore, for effective mitigation of single-beam SRS, the frequency difference between any two beamlets should be at the scale of $10^{-2}\omega_0$ \cite{Zhao2019}.

In the multibeam configuration, considering for the case of two pumps with an incident angle $\theta_v$ in vacuum, the common Langmuir wave shared by the monochromatic lasers at $n_e/n_c\leq\cos^4\theta_v/4$ is divided into multiple weakly coupled modes by the polychromatic lights with a broad spectrum $\Delta\omega_0\gtrsim3\%\omega_0$ \cite{zhao2021mitigation,2015Multibeam}. Meanwhile, the multibeam absolute region $n_e/n_c=\cos^4\theta_v(1-\delta\omega_0/\omega_0)/4$ is also shifted by the detuning. According to Eq. (5), the scattering lights of the polychromatic beams are detuned from each other when the frequency difference satisfies $\delta\omega_0\sim10^{-2}\omega_0$. Assume that each polychromatic light has $N$ beamlets, and every beamlet shares different common modes with all other polychromatic beamlets, i.e., there are $N^2$ backscattered common waves. The amplification coefficient for the polychromatic lights can be estimated as $\alpha_{p}=N^2\exp(1)$, and that for the monochromatic lasers with the same energy is $\alpha_{n}=\exp(N)$ based on $a_i=a_0/\sqrt{N}$. The amplification ratio $\alpha_{p}/\alpha_{n}$ is found to be less than one when $N>3$, and the energy fraction $\alpha_{p}^2/\alpha_{n}^2<1$ is always satisfied for the polychromatic beams. Therefore, polychromatic lights with $\Delta\omega_0\gtrsim3\%\omega_0$ and beamlet number $N\geq4$ can effectively control the intensity and saturation level of collective SRS in inhomogeneous plasmas.

Different from the convective instability, the degenerate modes, including absolute SRS and TPD, always develop in a local density range $[(\omega_i/2-a_ik_ic/4)^2,(\omega_i/2+a_ik_ic/4)^2]$ \cite{Zhao2019,zhao2020}, where the multiple beamlets share the common daughter waves. Therefore, the beamlets are decoupled from each other when their frequency differences satisfy
\begin{equation}
\delta\omega_0/\omega_0\gtrsim0.866a_i.
\end{equation}
The absolute instabilities are effectively mitigated when the decoupled beamlets dissatisfy the intensity threshold in inhomogeneous plasmas \cite{liu1974raman,siom1983}.

A scattering light produced by a certain color beam will be amplified again as a seed mode in the subsequent parametric excitation region where its frequency is equal to that of the scattering light developed by another color beam. The reamplification of scattering light in inhomogeneous plasmas is the major concern for the mitigation of SBS and cross-beam energy transfer (CBET) \cite{Zhao2019,bates2018mitigation}. The bandwidth of SBS scattering light can be estimated as $\Delta\omega_s\lesssim2\Delta(k_{pi}c_s)\sim0.004\omega_0$ with including the effect of plasma flow \cite{Zhao2019}, where $\Delta k_{pi}\sim2k_0<2\omega_0/c$ is the scale of ion-acoustic wavenumber, and $c_s\sim10^{-3}c$ is the ion-acoustic velocity. Therefore, the frequency difference $\delta\omega_0\gtrsim0.5\%\omega_0$ is sufficient to inhibit SBS and CBET \cite{marozas2018first}.

\section{Optical system design of the polychromatic driver}

In general, a polychromatic driver is able to generate numerous polychromatic beamlets with whole spectrum width larger than $40\%\omega_0$ \cite{cui-05,huang2021highly}. However, the actual color numbers should be determined according to the cost and complexity of the project, as well as the LPI mitigation effect. Here we propose a typical design example of polychromatic driver with total 16 colors of monochromatic beamlets, where the full spectrum width is $\Delta\omega_0=7.5\%\omega_0$, and the difference between every two adjacent frequencies is constant $\delta\omega_0=0.5\%\omega_0$. The four-color beamlets with adjacent frequency difference $\delta\omega_0=1\%\omega_0$ are synthesized to form a single polychromatic beam with a spectrum width $\Delta\omega_0=3\%\omega_0$. Therefore, each laser entrance flange is injected with a four-color beam, and the polychromatic lights of any two adjacent flanges have totally eight different color beamlets as shown in Fig. 2. Both the single-beam and collective LPIs can be effectively mitigated by the polychromatic lights under this configuration \cite{Zhao2019,Michel2009,Marozas}.

\begin{figure}
\centering
    \begin{tabular}{lc}
        \begin{overpic}[width=1\textwidth]{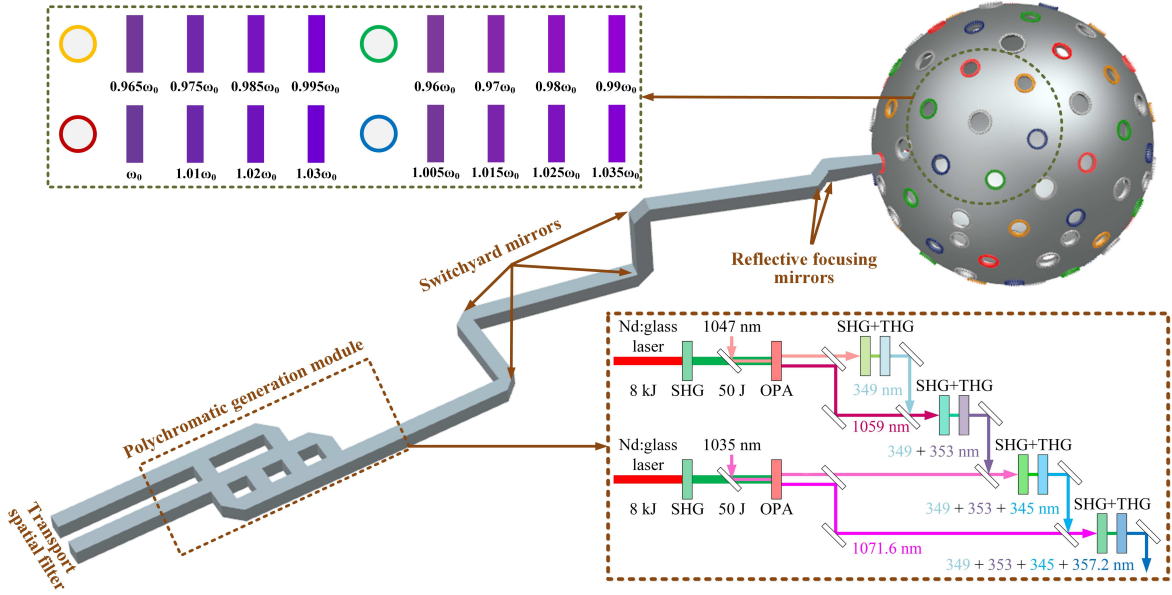}
        \end{overpic}
    \end{tabular}
\caption{Optical design diagram of the polychromatic driver. The polychromatic generation module is added between the transport spatial filter and the switchyard, and the detail optical design is indicated in the dotted box.
    }
\end{figure}

\begin{figure}
\centering
    \begin{tabular}{lc}
        \begin{overpic}[width=0.88\textwidth]{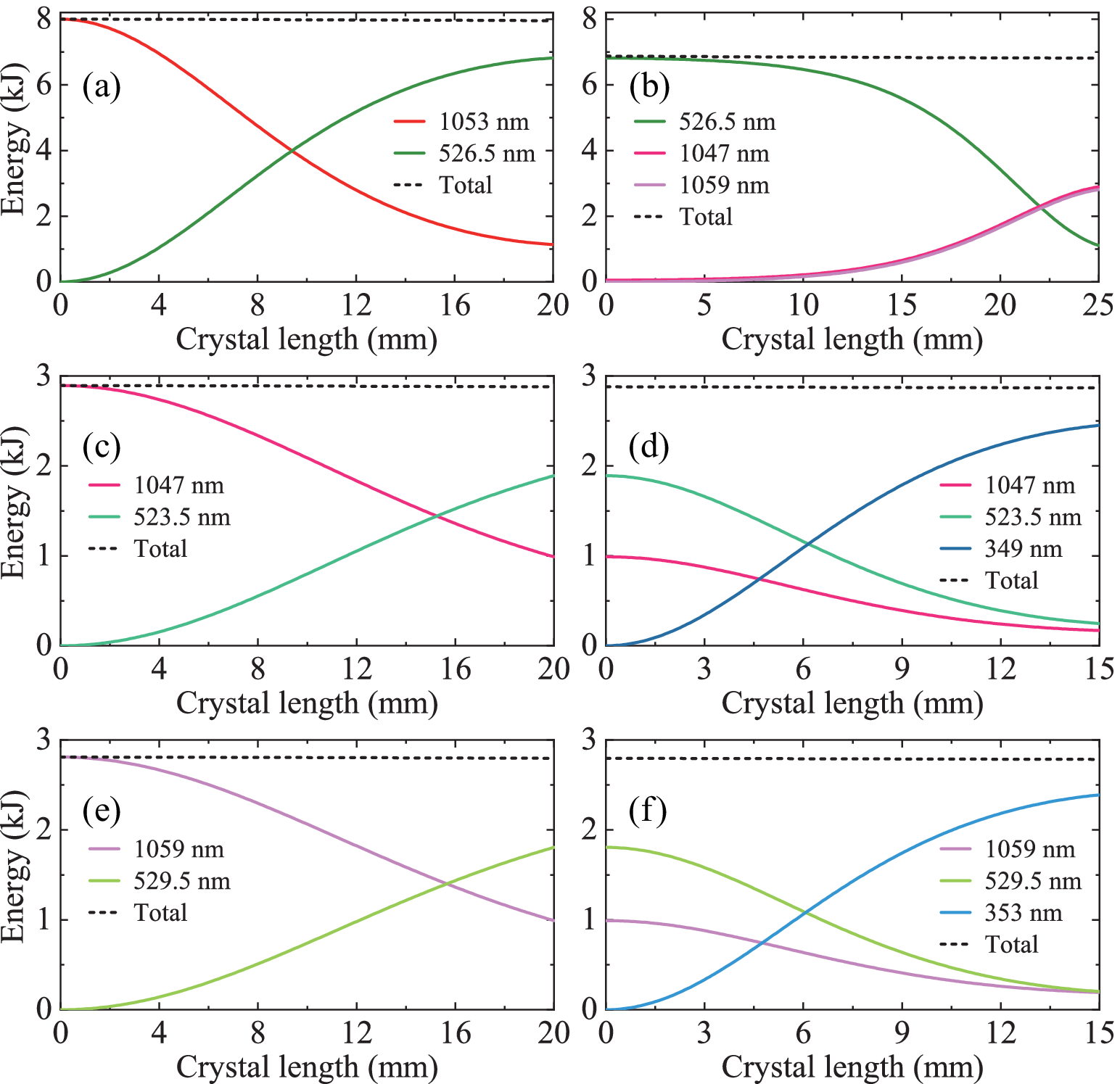}
        \end{overpic}
    \end{tabular}
\caption{Numerical simulations for the generation of polychromatic ultraviolet beam. The laser energies varied with the crystal length in the processes of (a)-(b) generation of the two-color light with 1047nm and 1059nm, (c)-(d) SHG and THG of the 1047nm laser, and (e)-(f) SHG and THG of the 1059nm laser, respectively.
    }
\end{figure}

Based on the polychromatic driver scheme, the duration of the multicolor beamlets is on the order of nanoseconds with a narrow bandwidth of each beamlet. Therefore, the type-I second-harmonic generation (SHG) process can be described by the following Eqs. (7)-(8) on the slowly varying envelope of plane waves \cite{cui-12,cui-13,cui-10}.
\begin{equation}
\frac{\partial A_1}{\partial x} = \frac{i\omega_1}{n_{1o} c} d_{\mathrm{eff}}A_2 A_1^*\exp(i \Delta k_\mathrm{I}x)
                                 + \frac{i}{2k_{1o}}\nabla_\perp^2A_1,
\end{equation}
\begin{equation}
\frac{\partial A_2}{\partial x} = \frac{i\omega_2}{2n_{2e} c} d_{\mathrm{eff}}A_1 A_1\exp(-i \Delta k_\mathrm{I}x)
                                 + \frac{i}{2k_{2e}}\nabla_\perp^2A_2
                                 + \tan\rho_{2e}\frac{\partial A_2}{\partial z},
\end{equation}
where $A$ is the complex amplitude of optical field, $n$ and $k$ respectively are the refractive index and wave vector in the nonlinear crystal, $\rho$ is the walk-off angle of the extraordinary light, $d_{\mathrm{eff}}$ is the effective nonlinear coefficient, and $\Delta k_\mathrm{I}$ is the phase mismatch $\Delta k_\mathrm{I}=k_{2e}-2k_{1o}$. The subscripts $o$ and $e$ represent the ordinary light and extraordinary light, and the subscripts $1$ and $2$ represent the fundamental wave and the corresponding second harmonic, respectively. The operator $\nabla_\perp^2=\partial^2_y+\partial^2_z$, where $y$ and $z$ are the transverse Cartesian coordinates in the plane orthogonal to the $x$ direction of propagation.

Type-II phase matching is used for the OPA and third-harmonic generation (THG) processes, and the corresponding nonlinear coupling equations are as follows
\begin{equation}
\frac{\partial A_1}{\partial x} = \frac{i\omega_1}{n_{1e}c}d_{\mathrm{eff}}A_3A_2^*\exp(i\Delta k_{\mathrm{II}}x)
                                 + \frac{i}{2k_{1e}}\nabla_\perp^2A_1
                                 + \tan\rho_{1e}\frac{\partial A_1}{\partial y},
\end{equation}
\begin{equation}
\frac{\partial A_2}{\partial x} = \frac{i\omega_2}{n_{2o}c}d_{\mathrm{eff}}A_3A_1^*\exp(i\Delta k_{\mathrm{II}}x)
                                 + \frac{i}{2k_{2o}}\nabla_\perp^2A_2,
\end{equation}
\begin{equation}
\frac{\partial A_3}{\partial x} = \frac{i\omega_3}{n_{3e}c}d_{\mathrm{eff}}A_1A_2\exp(-i\Delta k_{\mathrm{II}}x)
                                 + \frac{i}{2k_{3e}}\nabla_\perp^2 A_3
                                 + \tan\rho_{3e}\frac{\partial A_3}{\partial y},
\end{equation}
where $\Delta k_{\mathrm{II}}=k_{3e}-k_{2o}-k_{1e}$. The subscripts $1$, $2$, and $3$ refer to the fundamental, second and third harmonics in the THG process, and refer to the signal light, idler light and pump light in the OPA process, respectively. The initial parameter of optical field is input according to the energy, aperture and pulse duration of spatiotemporal shaping Nd:glass laser. The linear and nonlinear propagation terms of Eqs. (7)-(11) can be solved using the split-step Fourier method and the fourth order Runge-Kutta method, respectively \cite{cui-12,cui-10}.

The optical scheme for the efficient generation of high-power polychromatic laser is based upon the OPA and frequency conversion technologies indicated above, and an example for the generation of a four-color beam is exhibited in Fig. 3. The two-color light generator, including an Nd:glass laser (1053nm) and another signal light, is a unit of the polychromatic system. The second harmonic of Nd:glass laser (526.5nm) is the pump source to amplify the signal light and simultaneously produce the idler beam via OPA. The produced two-color beams with fundamental frequency are then separated due to their distinct polarizations, and each monochromatic beamlet is triple-frequency converted independently. The other two-color light is generated in the same way. Finally, the four-color beamlets are synthesized into a single beam with all the beamlets propagation in the same direction, and the polarization of each beamlet can be controlled by adding the plasma electrode Pockels cell \cite{cui-11}. The rms nonuniformity of kilojoule low-coherent lights with wavelength 529nm is around 40.7\%, which is acceptable for the optics \cite{gao2020high}. In the process of beam combination, reflective optical structures of the third-harmonic lights are utilized to reduce the damage.

In our case, the spatiotemporal shaping Nd:glass laser with an aperture of 350mm, 8kJ energy, and 3ns pulse duration is the main pulse \cite{cui-06,cui-07,cui-08,cui-09}. Two beams with wavelengths 1035nm and 1047nm are the signal lights, and both of them are pre-amplified to 50J via multistage OPA using the 526.5nm laser. As a result, four fundamental waves with wavelengths 1047nm, 1059nm, 1035nm and 1071.6nm are generated in the OPA process, and each of them is then frequency converted independently. The amplification and frequency-conversion processes of the 1047nm and 1059nm lasers using deuterated potassium dihydrogen phosphate (DKDP) are presented as examples in Figs. 3(a) and 3(b), where the refractive index, walk-off angle and effective nonlinear coefficient of DKDP crystal are obtained from Sellmeier equations and second-order nonlinear optical tensor \cite{cui-05,cui-12}. Note that DKDP is one of the optimal nonlinear crystals currently used in the high-power laser systems due to its high optical quality for kilojoule pulses \cite{dorrer2021broadband}. Meanwhile, DKDP is also used in the generation of low-coherent green laser at the energy scale of kilojoule owning to its broadband nature \cite{gao2020high}. The second harmonic of 1053nm laser with an energy of 6.8kJ is produced by a 20-mm-thick DKDP crystal, which is then used to amplify the 1047nm laser via the OPA process. As a result, the 1047nm laser is amplified to 2.9kJ though a 25-mm-thick DKDP crystal, and a 1059nm laser with an energy of $\sim2.85$kJ is generated in the meanwhile.

In order to obtain the third-harmonic light efficiently, the efficiency of the SHG should be around 66.7\%, so that the photon number fraction of the fundamental wave and the second-harmonic wave is 1:1, and therefore the corresponding energy ratio is 1:2 for efficient conversion \cite{cui-10}. Under our design scheme, 20-mm- and 15-mm-thick DKDP crystals are used for the SHG and THG of 1047nm laser, respectively, as indicated in Figs. 3(c) and 3(d). The other two DKDP crystals with almost the same thickness are used for the SHG and THG of 1059nm laser, as displayed in Figs. 3(e) and 3(f). The obtained two-color beamlets with wavelengths 349nm and 353nm have an equal energy of $\sim2.4$kJ. Therefore, the total energy conversion efficiency from the Nd:glass laser to polychromatic ultraviolet light is about 60\%. The generation, amplification, and frequency conversion mechanisms for the lasers with wavelengths of 1035nm and 1071.6nm are the same as the above examples. The polychromatic beamlets are synthesized and focused via the dichroic mirror and off-axis parabolic mirror, respectively \cite{stolz2018transport}. Both of the optics are widely used in the high-power laser systems due to their relatively high damage threshold. Finally, one obtains a polychromatic ultraviolet beam with total energy $\sim9.6$kJ, 3ns pulse duration, and four different wavelengths 345nm, 349nm, 353nm, and 357.2nm, where the difference of two adjacent frequencies is $\delta\omega_0\approx1.1\%\omega_0$ regarding to the central wavelength 351nm.

\begin{figure}
\centering
    \begin{tabular}{lc}
        \begin{overpic}[width=1\textwidth]{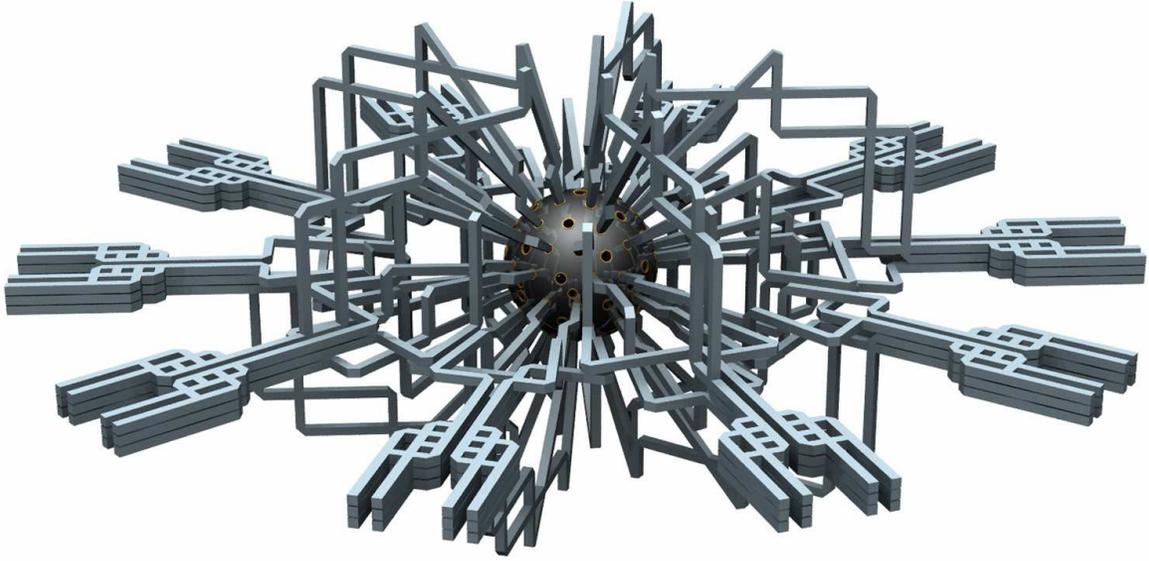}
        \end{overpic}
    \end{tabular}
\caption{Schematic diagram of the optical paths and switchyard in target area of the polychromatic driver with 60 incident beams as example.
    }
\end{figure}

A polychromatic direct-drive system with 60 incident beams is presented as an example in Fig. 4, which describes the overall structure of the polychromatic light generation, transmission and focusing based upon the single beam diagram in Fig. 2. The equal optical paths of all the laser beamlines are realized by employing the baseline algorithm and symmetrical operations \cite{ren2015,Ren2017}, and the switchyard in the target area is displayed in Fig. 4. The optical units used in the processes of reflection and focusing are reflective to reduce the optic component damage. The third-harmonic beams injected into the target chamber are obtained by filtering the fundamental and second-harmonic frequency components with the multiple color separation mirrors. Note that the basic optical technologies and designs are also applicable to the indirect-drive scheme.

Two key refitments are sufficient to upgrade the existing facility to generate polychromatic lights as shown in Fig. 2 \cite{spaeth}, which introduce moderate changes into the whole drive configuration compared to the conventional broadband system \cite{eimerl2016stardriver}. First, the polychromatic generation module is installed in the upstream of the switchyard. Second, the focusing lens in the final optics assembly are replaced by an off-axis parabolic mirror and a projecting mirror. Regarding the megajoule facilities such as NIF \cite{spaeth} and Laser Megajoule \cite{Miquel2018}, the wavelength detunings in a quad can effectively suppress the collective instability modes by adding polychromatic generation modules.

In a brief conclusion, the amplification and efficient frequency conversion of the broadband Nd:glass laser are the major obstacles to obtaining high-power broadband ultraviolet lasers. The conversion efficiency of a fundamental laser to the third-harmonic laser with bandwidth $1.5\%\omega_0$ as reported is approximately 20\% \cite{cui-04}, which is only a third of the efficiency of the polychromatic light with a spectrum width $\Delta\omega_0=3\%\omega_0$. Moreover, the refitment of the broadband laser driver is extremely complicated and incurs enormous cost owing to the upgrade of the front end, preamplifier, and main amplifier. Therefore, the proposed polychromatic driver is a feasible and effective way to mitigate the LPIs.

\section{Particle-in-cell simulations of LPI mitigation via polychromatic lights}

\subsection{Suppression of LPIs with polychromatic lights under the indirect-drive scheme}

\begin{figure}
\centering
    \begin{tabular}{lc}
        \begin{overpic}[width=1\textwidth]{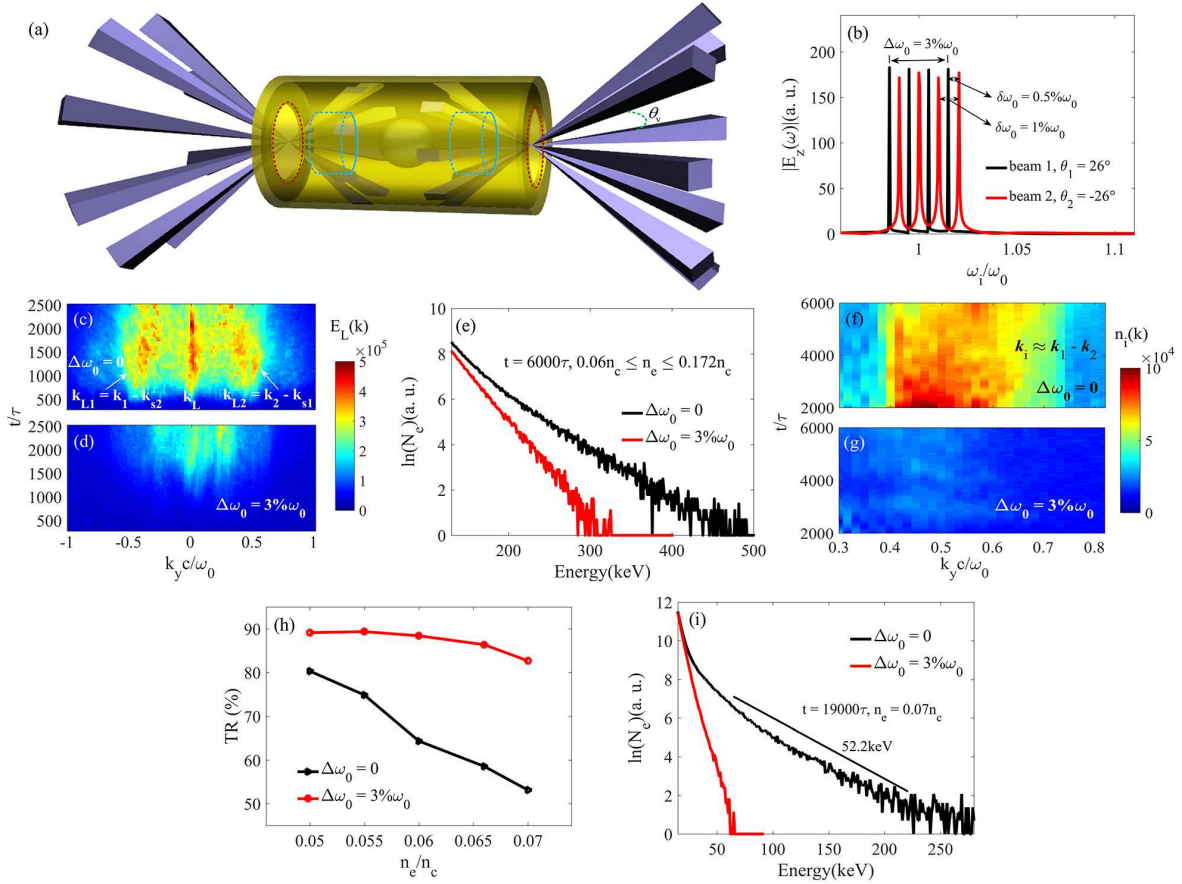}
        \end{overpic}
    \end{tabular}
\caption{Comparison of simulation results of LPIs developed by monochromatic lasers and polychromatic lights under the indirect-drive scheme. (a) Configuration of indirect drive with LPI regions considered in the simulations. (b) Spectra of the two incident polychromatic beams in the 2D simulations. (c)-(g) Two-beam simulation results correspond to the region labeled by the red dotted line in (a). Transverse wavenumber evolutions of the electrostatic field developed by (c) monochromatic lasers $I_0=1.1\times10^{15}$W/cm$^2$ and (d) polychromatic lights with each $I_i=2.75\times10^{14}$W/cm$^2$, where the longitudinal wavenumber is summated from $k_xc=1.1\omega_0$ to $k_xc=1.45\omega_0$. (e) Comparison of electron hot tail developed by monochromatic lasers and polychromatic lights at $t=6000\tau$. $N_e$ is the relative number of electrons. Transverse wavenumber evolutions of ion acoustic waves stimulated by the crossing of (f) monochromatic lasers and (g) polychromatic lights, where the longitudinal wavenumber is summated from $k_xc=-0.05\omega_0$ to $k_xc=0.05\omega_0$. (h)-(i) 1D simulation results of the single beam interaction with the filled gas in hohlraum correspond to the region marked by the blue dotted line in (a). (h) Transmission rate of incident energies at different plasma densities. (i) Electron energy distributions heated by monochromatic laser and polychromatic light at $t=15000\tau$ under $n_e=0.07n_c$.
    }
\end{figure}

All the PIC simulations have been carried out using the OSIRIS code \cite{fonseca2002osiris,hemker2015particle}. Soft X-rays, generated by laser irradiation on the hohlraum wall, compress the target under the indirect-drive configuration. To demonstrate the suppression effects of LPIs with polychromatic lights, two dominant parameter intervals of laser plasma interactions are investigated here including the plasma at the entrance hole and the ionized gas in hohlraum as shown in Fig. 5(a). We first consider the inhomogeneous plasma at the entrance hole of the hohlraum \cite{2015Multibeam,dewald2016generation}. The third-harmonic laser intensity is $I_0=1.1\times10^{15}$W/cm$^2$. Under the same incident energy, each beamlet intensity of the polychromatic light is $I_i=2.75\times10^{14}$W/cm$^2$ at beamlet number $N=4$ and spectrum width $\Delta\omega_0=3\%\omega_0$. Figure 5(b) presents the spectra of the two polychromatic lights, where the black and red lines respectively correspond to the two pump beams with incident angles $\theta_1=26^\circ$ and $\theta_2=-26^\circ$, i.e., the angle between the two beams in vacuum is $\theta_v=\theta_1-\theta_2=52^\circ$. Each polychromatic light has four-color beamlets with whole spectrum width $\Delta\omega_0=3\%\omega_0$, and the difference between every two adjacent frequencies of the two beams is $\delta\omega_0=0.5\%\omega_0$ according to Fig. 5(b).

The lasers are incident from the two transverse boundaries in the two-dimensional (2D) simulations for the stimulated scattering instabilities. The coupling term of incident light and scattering light is $\vec{a}_0\cdot\vec{A}_s=a_0A_s\cos(\phi)$, where $\phi$ is the angle of polarizations. It should be noted that $\cos(\phi)=1$ is always satisfied for the s-polarized (the electric field of light is perpendicular to the plane) incident beams. Therefore, s-polarized beams are used in our 2D simulations for the stimulated scattering instabilities, which normally lead to relatively larger growth rates than p-polarized (the electric field of light is parallel to the plane) ones \cite{zhao2021mitigation}. In our 2D simulations, the longitudinal plasma density profile is $n_e/n_c=0.06(1+x/500)$ with the density range of $0.06n_c\leq n_e\leq0.172n_c$. The transverse width of the plasma is 44$\lambda$, and the resolutions are $dx/\lambda=dy/\lambda=0.0286$ with 25 particles per cell. The ion mass is $m_i=3672m_e$ with an effective charge $Z=1$ for all the simulations. The electron and ion temperatures are $T_e=2$keV and $T_i=0.8$keV, respectively. The temporal evolutions of electrostatic fields developed by multibeam SRS are exhibited in Figs. 5(c) and 5(d), which are obtained by summating the wavenumber distributions along the longitudinal direction $1.1\omega_0\leq k_xc\leq1.45\omega_0$. Multibeam SRS has been strongly developed at $t=500\tau$ for the monochromatic lasers on the basis of Fig. 5(c), where $\tau$ is the light period. The sidebands are driven by the beating of a pump with the scattering light of another one, i.e., the transverse distributions of $\vec{k}_{L1}=\vec{k}_1-\vec{k}_{s2}$ and $\vec{k}_{L2}=\vec{k}_2-\vec{k}_{s1}$. With respect to polychromatic lights, only weak SRS signals can be found later at $t=1500\tau$ according to Fig. 5(d). The reflectivities of both SRS and SBS are reduced by a factor of more than three using polychromatic beams, and the mitigation of the electron hot tail is displayed in Fig. 5(e). The cross beams transfer energies through the stimulated ion acoustic waves, where one of the pump beams is a seed light to develop SBS via beating with the other one. Temporal evolutions of the CBET ion mode developed by different types of incident lights are displayed in Figs. 5(f) and 5(g). An intense CBET mode can be observed for monochromatic lasers near $\vec{k}_{pi}\approx\vec{k}_1-\vec{k}_2$, i.e., $k_{y}c\approx0.6\omega_0$ and $k_{x}c\approx0$ in Fig. 6(f). However, the instability has not been clearly developed in the case of polychromatic beams according to Fig. 6(g). The transmission efficiency of the incident energy is enhanced from 80.4\% to 91.4\% by polychromatic lights. Therefore, polychromatic drivers can save at least 11\% of the incident energy from the stimulated scattering instabilities at the entrance hole.

Different from the plasma at the entrance hole, the gas filled in hohlraum is almost homogeneous, where the electron density is always less than $0.07n_c$ \cite{le2014observation,hall2017relationship}. According to Eq. (3), the decoupling threshold of SRS is found to be around $0.75\%\omega_0$ at $n_e=0.07n_c$ under the beamlet amplitude $a_i=0.005$. Therefore, the beamlets of the polychromatic light are decoupled from each other. In order to investigate LPI development inside the hohlraum, comparative one-dimensional (1D) simulations for monochromatic laser and polychromatic light interaction with different density plasmas have been carried out as shown in Figs. 5(h)-5(i). A laser pulse with a plateau of 15000$\tau$ and two edges of 25$\tau$ at the front and end is incident from the left boundary of the simulation box. We have taken 100 cells per wavelength and 100 particles per cell. Large-scale plasmas with a whole length of $3000\lambda$ are set in the simulations to include the convective effects. The electron and ion temperatures are $T_e=2.5$keV and $T_i=1$keV, respectively. Figure 5(h) indicates that the transmission efficiency of the incident energy varies with the plasma density. Polychromatic beam can save at least 8.8\% of the incident energy from LPIs compared with the monochromatic laser at $n_e=0.05n_c$. The energy-saving ratio of polychromatic light increases with the plasma density, which is up to 29.57\% at $n_e=0.07n_c$. Electron energy distributions developed by different pump beams are displayed in Fig. 5(i). A saturated hot electron tail with $T_e=52.2$keV can be found for the monochromatic laser at $t=19000\tau$ (i.e., 22.3ps). However, the electrons have not been obviously heated at this time for the polychromatic light.

In a brief conclusion, the laser energy loss and hot-electron generation in the entrance hole and hohlraum are effectively mitigated by the polychromatic lights, where the whole transmission efficiency is enhanced by more than 40\% from the high-density gas filled target. Therefore, compared to the trade-off scheme of near-vacuum hohlraum, both the energy loss and the dynamic behavior of gas filled hohlraum can be controlled using the polychromatic lights, which may improve the system stability.

\subsection{Suppression of LPIs with polychromatic lights under the direct-drive scheme}

\begin{figure}
\centering
    \begin{tabular}{lc}
        \begin{overpic}[width=0.88\textwidth]{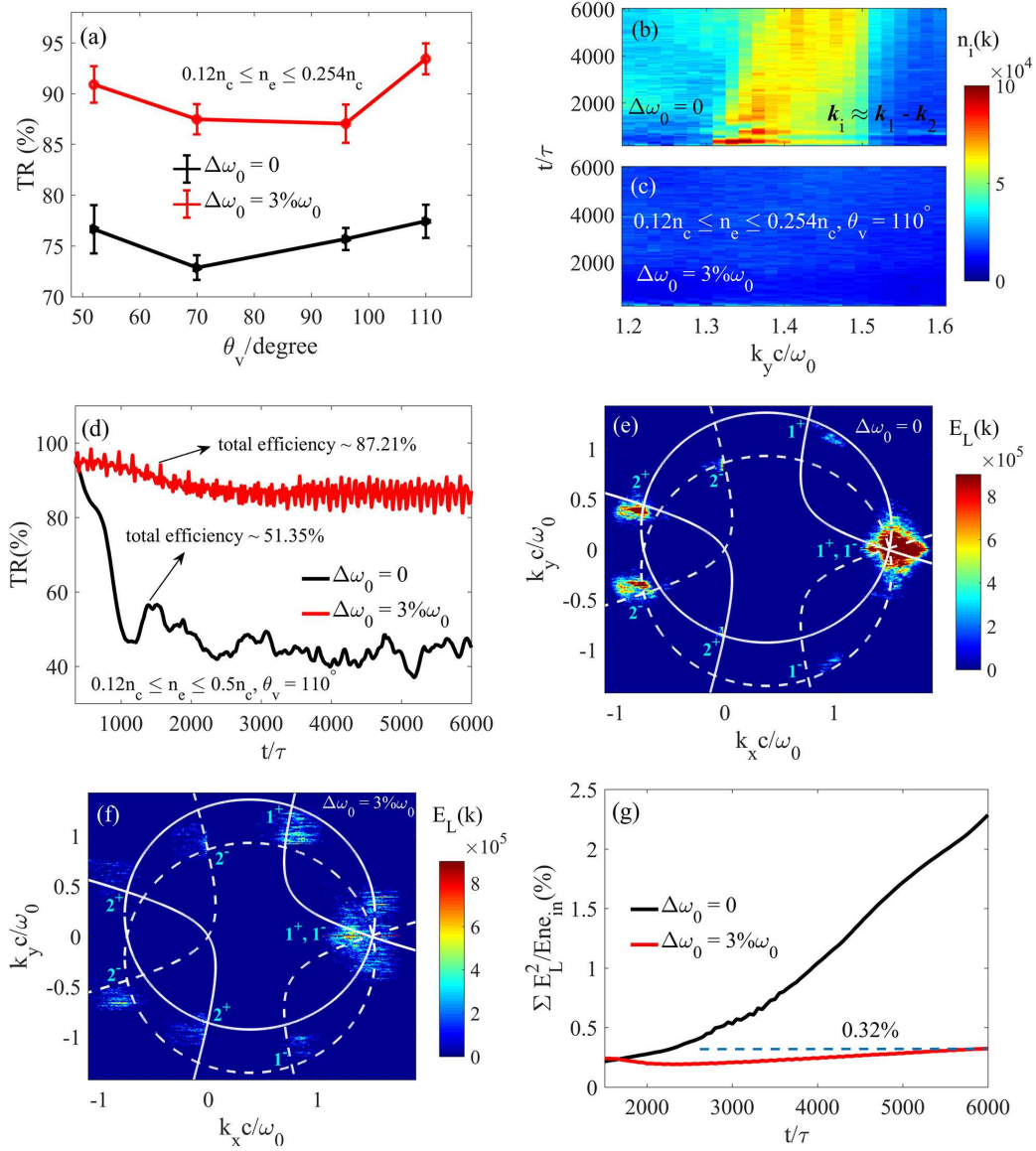}
        \end{overpic}
    \end{tabular}
\caption{Comparison of 2D simulation results of LPIs developed by monochromatic lasers and polychromatic lights under the direct-drive scheme. (a)-(d) Simulation results of stimulated scattering instabilities with s-polarized incident beams. (a) Transmission rate of incident energies for different incident angles with density range of $0.12n_c\leq n_e\leq0.254n_c$. (b)-(c) Simulation example at $\theta_v=110^\circ$ of (a). Transverse wavenumber evolutions of ion acoustic waves stimulated by the crossing of (b) monochromatic lasers and (c) polychromatic lights, where the longitudinal wavenumber is summated from $k_xc=-0.05\omega_0$ to $k_xc=0.05\omega_0$. (d) Transient transmission efficiencies of incident lights in a larger scale plasma with $0.12n_c\leq n_e\leq0.5n_c$ and $\theta_v=110^\circ$. (e)-(g) Comparative results for multibeam TPD at plasma density $n_e\sim0.232n_c$. The incident angle between the two p-polarized beams is $\theta_p=59^\circ$ in plasma. Wavenumber distributions of electrostatic field summarized from $t=1000\tau$ to $t=6000\tau$ for (e) monochromatic lasers and (f) polychromatic lights. The intersections of the white curves show the most intense TPD growing regions in the wave-vector space. (g) Temporal evolutions of the fraction of incident laser energy $\mathrm{Ene}_{in}=\Sigma|E_z(t)|^2$ converted into electrostatic energy $\Sigma(|E_L(x,y)|\omega_L/\omega_0)^2$.
    }
\end{figure}

Lasers directly irradiate on the target under the direct-drive scheme, and interact with the large scale plasma formed by the ablation. Figures 6(a)-6(d) present the comparative simulation results of stimulated scattering instabilities, where the longitudinal plasma density profile is $n_e/n_c=0.12\exp(x/1000)$ with the density range $0.12n_c\leq n_e\leq0.254n_c$, and the incident beams are s-polarized. The ion flow velocity assumed to be linear as $v_f/c=-0.0035+6.84\times10^{-4}x/1000$. The plasma temperatures are the same as those in the 2D simulations in Fig. 5. Figure 6(a) shows the transmission rate of the incident energy under different angles between the two beams, which indicates that polychromatic beams can save at least 10\% of the incident energy from LPIs in the SRS developing regime. As an example, the SRS reflectivity at $\theta_v=110^\circ$ is reduced from 5.1\% to 1.83\% by polychromatic lights. The inhibition of SRS weakens the electrostatic field, and therefore reduces hot electron generation. The hot electrons with kinetic energies $\geq50$keV developed by polychromatic lights are greatly reduced compared to monochromatic lasers, where the total energy of the former is only 6.7\% of that of the latter. According to the temporal evolutions of the CBET ion mode shown in Figs. 6(b) and 6(c), the intense CBET mode $k_{y}c\approx1.5\omega_0$ driven by monochromatic lights is significantly reduced by polychromatic lights. To investigate the mitigation of stimulated scattering instabilities at a wide density range by polychromatic lights, a large scale plasma with density region $0.12n_c\leq n_e\leq0.5n_c$ is considered in the simulation example at $\theta_v=110^\circ$, and the result is shown in Fig. 6(d). The transient transmission efficiency is reduced marginally at $t=2000\tau$ for the polychromatic lights. However, a dramatic reduction in the transient transmission efficiency is found at $t=1000\tau$ for the monochromatic lasers. In addition, the total efficiency of the energy transfer has been enhanced by 35.86\% with using polychromatic lights. Amplitude modulation is a universal phenomenon of broadband lasers. However, the modulation time scale is much shorter than the fluid scale, and the amplitude modulation just introduces a weak fluctuation into the saturation level when the beamlets are weakly coupled as indicated in Fig. 6(d). An efficient smoothing effect has been found for the low-coherence light, where the rms nonuniformity is reduced by around three times and the smoothed focal spot has a shorter longitudinal speckle length than normal monochromatic lasers \cite{gao2020high}.

Unlike the stimulated scattering instabilities, the TPD instability is the decay of a laser into two equal-frequency Langmuir waves near $0.25n_c$. Therefore, we only consider the density region $0.225n_c\leq n_e\leq0.242n_c$ with longitudinal profile $n_e/n_c=0.225\exp(x/1000)$ in the simulations. Each beamlet amplitude of the polychromatic light in our simulations is $a_i=0.005$, which satisfies the decoupling threshold for TPD $0.433\%\omega_0<1\%\omega_0$ according to Eq. (6). The transverse width of the plasmas is 94$\lambda$. The electron and ion temperatures are $T_e=2$keV and $T_i=1$keV, respectively. The angle between the two incident beams is $\theta_p=59^\circ$ in the plasma. The other laser parameters are unchanged. Compared with the monochromatic lasers, the instability regions of the collective TPD modes are broadened and weakened by the polychromatic lights according to Figs. 6(e) and 6(f), where the intersections of the white curves show the most intense TPD growing regions in the wave-vector space \cite{Vu2010The}. The submodes of the electrostatic fields developed by different color beamlets are weakly coupled in the phase space \cite{zhao2021mitigation}. Therefore, the total saturation level is reduced by the wavenumber detuning. The fraction of incident laser energy converted into electrostatic energy can directly reflect the TPD development as displayed in Fig. 6(g). The maximum energy fraction is approximately 0.32\% at $t=6000\tau$ for the polychromatic lights. However, the electrostatic energy developed by the monochromatic lasers increases with time without saturation even at $t=6000\tau$, and the peak is about eight times higher than that of polychromatic lights. The total energy of the hot electrons with kinetic energies $\geq50$keV is reduced by 86.3\% with using polychromatic lights.

As a whole, polychromatic lights can save more than 35\% of incident energy from LPI scattering and TPD under the direct-drive scheme. The total energy of the hot electrons with kinetic energies $\geq50$keV is reduced by more than 86\% with using polychromatic lights. The mitigation of incident energy loss, CBET and hot electron generation can effectively enhance the compression efficiency of target.

\section{Summary}

We have proposed a design of high-power polychromatic ultraviolet laser drivers for inertial fusion energy in order to enhance the beam-target coupling. A polychromatic beam is composed of different-color monochromatic beamlets with a whole spectrum width of $3\%\omega_0$, which is generated based upon the OPA and frequency conversion technologies. Each monochromatic beamlet is frequency converted independently, which skirts the low conversion efficiency of the broadband triple frequency, a major obstacle to obtaining high-power broadband ultraviolet lasers. The project cost and construction difficulty of the proposed scheme are well controlled, which introduce moderate changes into the whole drive configuration.

We have evaluated the LPI inhibition effects with such a laser driver via PIC simulations. It is found that the polychromatic lights not only delay the growth of LPIs, but also significantly reduce the LPI extent at similar times of observation. A clear enhancement of the laser transmission efficiency and hot electron reduction is found with the polychromatic lights both for indirect-drive and direct drive schemes. The single-beam and collective LPIs, such as SRS, TPD, and CBET, are all effectively suppressed by the wavenumber detuning of the polychromatic beamlets, which may save up to 35\% of the incident energy from LPIs. Our simulation results confirm the significant potential of polychromatic beams to enhance the laser-target coupling efficiency for inertial fusion energy.

\section{Acknowledgements}

This work was supported by the National Natural Science Foundation of China (Nos. 12005287, 12004404 and 12135009), the Strategic Priority Research Program of Chinese Academy of Sciences (Grant Nos. XDA25050800 and XDA25050100), and the Natural Science Foundation of Shanghai (Nos. 19YF1453200 and 18YF1425900). The authors would like to acknowledge the OSIRIS Consortium, consisting of UCLA and IST (Lisbon, Portugal) for providing access to the OSIRIS 4.0 framework.

\end{document}